\documentclass[12pt,a4paper]{revtex4}
\usepackage{amsmath,amssymb,graphics,epsfig,subfigure}
\usepackage{color}
\usepackage[colorlinks,linkcolor=red,anchorcolor=red,citecolor=green]{hyperref}
\usepackage{geometry}
\begin{document}

\thispagestyle{empty}

\begin{center}

\title{A new measure of thermal micro-behavior for the AdS black hole}

\author{Zhen-Ming Xu$^{1,2,3,4,}$\footnote{E-mail: xuzhenm@nwu.edu.cn}, Bin Wu$^{1,2,3,4,}$\footnote{E-mail: binwu@nwu.edu.cn}, Tao Yang$^{1,2,3,4,}$\footnote{E-mail: yangt@nwu.edu.cn (Corresponding author)}
        and Wen-Li Yang$^{1,2,3,4,}$\footnote{E-mail: wlyang@nwu.edu.cn}
        \vspace{5pt}\\}

\affiliation{$^{1}$Institute of Modern Physics, Northwest University, Xi'an 710127, China\\
$^{2}$School of Physics, Northwest University, Xi'an 710127, China\\
$^{3}$Shaanxi Key Laboratory for Theoretical Physics Frontiers, Xi'an 710127, China\\
$^{4}$Peng Huanwu Center for Fundamental Theory, Xi'an 710127, China}

\begin{abstract}
Inspired by the hypothesis of the black hole molecule, with the help of the Hawking temperature, entropy and the thermodynamics curvature of the black hole, we propose a new measure of the relation between the interaction and the thermal motion of molecules of the AdS black hole as a preliminary and coarse-grained description. The measure enables us to introduce a dimensionless ratio to characterize this relation and show  that there is indeed competition between the interaction among black hole molecules and their thermal motion. For the charged AdS black hole, below the critical dimensionless pressure, there are three transitions between the interaction state and the thermal motion state. While above the critical dimensionless pressure, there is only one transition. For the Schwarzschild-AdS black hole and five-dimensional Gauss-Bonnet AdS black hole, there is always a transition between the interaction state and the thermal motion state.
\end{abstract}

\maketitle
\end{center}


\section{Motivation}
The exploration of microscopic properties of the black hole is a hot issue in theoretical physics. Traditional general relativity holds that there is no matter structure inside a black hole, except for the singularity. However, the proposal of the black hole temperature and area entropy provides a new perspective to re-understand properties of the black hole~\cite{Hawking1975,Bekenstein1973,Bardeen1973}. With the development of the black hole thermodynamics~\cite{Hawking1983,Chamblin1999,Wald2001,Padmanabhan2010,Carlip2014}, especially the introduction of the extended phase space~\cite{Kastor2009}, black holes show abundant critical behaviors~\cite{Dolan2011a,Kubiznak2012,Spallucci2013,Kubiznak2017}, which indicates that the black hole has some unknown microscopic characteristics.

Recently, based on the Ruppeiner thermodynamic geometry~\cite{Ruppeiner1995}, a new abstract concept ``black hole molecule'' has been proposed in Ref.~\cite{Wei2015}, which provides a new way to study the micro behavior of the black hole roughly and phenomenologically. They think that the interior of a black hole is a kind of fluid, and the fluid composed of these black hole molecules can give the microstructure of the black hole. This is similar to the atomic hypothesis put forward by Boltzmann more than a century ago. Although we don't know exactly what the black hole molecules are, as a preliminary exploration of the black hole microstructure, this abstract concept is very useful for understanding some micro properties of black holes. Among them, the role of the Ruppeiner geometry is particularly important. It introduces a thermodynamic metric to represent the thermodynamic fluctuation theory. Meanwhile, the components of the inverse thermodynamic metric correspond to second moments in the fluctuation theory of equilibrium thermodynamics. The original form of the thermodynamic metric is the second derivative of entropy to other generalized coordinates (or other thermodynamic quantities).

In the process of studying the Ruppeiner geometry of black holes, the thermodynamic curvature is the most important physical quantity. It has two important roles:
\begin{itemize}
  \item {\em Analyzing the phase transition} ~It is believed that the divergence of the thermodynamics curvature should also correspond to some kind of the phase transition of the system. There are two ways of understanding the divergence. On the one hand, the divergence of the thermodynamic curvature corresponds to the configuration of the extreme black hole, or perhaps, the black hole/extremal black hole transition~\cite{Wei2015,Cai1999,Mirza2007,Dehyadegari2017,Miao2018,Zangeneh2018,Yazdikarimi2019,Guo2019,Miao2019,Xu2019,Xu2020,Xu2020b,Ghosh2020}. On the other hand, the divergence of the thermodynamic curvature corresponds to the divergence of the heat capacity, i.e., a second-order phase transition~\cite{Shen2007,Liu2010,Niu2012,Wang2020,Mansoori2015,Mansoori2019,Quevedo2008,Bhattacharya2017,Bhattacharya2020,Zhang2015}. However, at present, there is no theoretical requirement that the divergent point of the thermodynamic curvature must correspond to the divergence of the heat capacity or the configuration of the extreme black hole. No matter what the divergent point of the thermodynamic curvature corresponds to, it all can be regarded as a reasonable and feasible way to understand the divergent behavior of the thermodynamic curvature.
  \item {\em Discussing the interaction in a thermodynamic system} ~The thermodynamic curvature is just as a new physical quantity. We only care about the role of the thermodynamic curvature itself in the microscopic behavior of the system. For some statistical mechanics models that we are already familiar with, there is an empirical but still a hypothetical observation in the view of thermodynamics geometry, that is the thermodynamic curvature can have a corresponding relationship with the interaction among the constituent molecules of the system ~\cite{Ruppeiner2010,Ruppeiner2014}. However, for the black hole system, the situation is slightly different and we need to explain the logic about exploring its micro behavior. For the statistical mechanics, general lore is that if we know the microscopic dynamics of a system, its thermodynamical properties could be derived from the statistic physics of the system. While the inverse process does not hold in general, namely we cannot know the micro-dynamics of the system from its thermodynamics. Turning to the black hole thermodynamic system, at present, its thermodynamic properties have been widely discussed, but the micro-dynamics behavior is still in the exploration stage due to the lack of the theory of quantum gravity. Therefore, if we want to explore some micro behaviors of black holes, we need to make some appropriate assumptions. Comparing with the mature research system of the usual statistical mechanical models, we can preliminarily think that the above mentioned inverse process may be applied to the black hole system. Coincidentally, the idea of this inverse process is reflected in the exploration of the microscopic behavior of black holes by the Ruppeiner thermodynamic geometry. Although some issue discussed in this scheme remains debatable within the community, it seems that there is no more suitable method than this one from the thermodynamic point of view. With the help of the abstract concept of the black hole molecule, through analogy, we can think that there is an interaction among the molecules that make up the black hole, and the above empirical observation also applies to black holes. In Ref.~\cite{Ruppeiner2008}, Ruppeiner gives the physical meaning of the thermodynamic curvature in the black hole system. Afterward, Refs.~\cite{Miao2018,Wei2019a} further point out that the absolute value of the thermodynamic curvature measures the strength of the interactions among black hole molecules phenomenologically or qualitatively. Recently, for a class of black hole systems with zero heat capacity at constant volume, there are two ways to deal with its thermodynamic curvature. One is to introduce the normalized thermodynamic curvature by treating the heat capacity at constant volume as a constant very close to zero~\cite{Wei2019a,Wei2019b,Wei2019c}. The other is to regard the entropy in the thermodynamic metric is a function of the mass (i.e. enthalpy) and other thermodynamic quantities of the AdS black holes~\cite{Xu2019,Xu2020,Xu2020b,Ghosh2020}. Both of these schemes can well analyze the interaction among molecules of the black hole system.
\end{itemize}

Hence one can see that the thermodynamics geometry can be considered as one of the important methods to explore the micro-information of black holes completely from the perspective of thermodynamics. In the past, people only used the positive/negative of the thermodynamic curvature to analyze the type of the interaction among black hole molecules in the black hole thermodynamic system, and used the magnitude of the thermodynamic curvature to qualitatively describe the strength of the interaction. We know that at the micro level, the interaction among the molecules that make up the system always competes with the thermal motion of the molecules themselves. Hence for the black hole thermodynamics system, this competition should exist. How do we describe it? This is the main motive of our present work.

Under the hypothesis of black hole molecules, the product of the temperature and entropy of the black hole can reflect the thermal motion of black hole molecules in terms of the theory of the molecular thermal motion. At the same time, in view of the idea of the thermodynamics geometry, we can approximately think that the thermodynamic curvature can be regarded as a measure of the interaction among black hole molecules. In our (three authors of this paper) previous work~\cite{Xu2019}, in the analysis of the thermal micro behavior of the Reissner-Nordstr\"{o}m  black hole, we give a signal about the relation between the thermal motion of black hole molecules and their interaction. Therefore in this paper, with the help of these two important tools, we try to make a preliminary and coarse-grained description of the relation between the interaction among black hole molecules and their thermal motion in the AdS background, so that we can have a new understanding of the possible microscopic behavior of black holes. Taking the AdS black holes as examples, we give a preliminary description of the relation between the interaction among black hole molecules and their thermal motion, and find the transition point between the state dominated by the interaction and the state dominated by the thermal motion in the black hole system, which also reveals the mechanism of black hole phase transition from the micro level completely from the perspective of thermodynamics.

The paper is organized as follows. In section \ref{sec2}, we give a brief introduction of the Ruppeiner thermodynamic geometry. In section \ref{sec3}, we introduce a dimensionless ratio to characterize the relation between the interaction and the thermal motion of molecules of the AdS black hole. Finally, we devote to drawing our conclusion in section \ref{sec4}.

\section{Ruppeiner geometry}\label{sec2}
Now we are going to provide a brief introduction of the Ruppeiner thermodynamic geometry, which originates from the fluctuation theory of the equilibrium thermodynamics~\cite{Ruppeiner1995,Ruppeiner2008}. Consider an equilibrium isolated thermodynamic system with total entropy $S$, and divide it into a small subsystem $S_B$ and a large subsystem $S_E$. We additionally require that $S_B \ll S_E \sim S$. Then the total entropy of the system reads as
\begin{equation}
    S(x^0,x^1,\cdots)=S_B(x^0,x^1,\cdots)+S_E(x^0,x^1,\cdots),
\end{equation}
where the parameters $x^0$, $x^1$, $\cdots$ stand for the independent thermodynamic variables. For a system in equilibrium state, the entropy $S$ has a local maximum value $S_0$ at $x_0^\mu$ ($\mu=0,1,2,\cdots$). Hence at the vicinity of the local maximum, we have
\begin{equation}
    S=S_0+\frac{\partial S_B}{\partial x_B^\mu}\Delta x^\mu_B
          +\frac{\partial S_E}{\partial x_E^\mu}\Delta x^\mu_E
      +\frac{1}{2}\frac{\partial^2 S_B}{\partial x_B^\mu \partial x_B^\nu}\Delta x^\mu_B \Delta x^\nu_B
       +\frac{1}{2}\frac{\partial^2 S_E}{\partial x_E^\mu \partial x_E^\nu}\Delta x^\mu_E \Delta x^\nu_E
       +\cdots.
\end{equation}
The first derivative terms in the above equation cancel each other due the conservation of the entropy of the equilibrium isolated system under the virtual change. Compared with the second derivative term of $S_B$, the one of $S_E$ can be ignored because $S_E$ is of the same order as that of the whole system ($S_E \sim S$). Finally we promptly arrive at
\begin{equation}
\Delta S =S_0-S \approx -\frac{1}{2}\frac{\partial^2 S_B}{\partial x_B^\mu \partial x_B^\nu}\Delta x^\mu_B \Delta x^\nu_B.
\end{equation}
According to the fluctuation probability given by Einstein’s formula $P\propto e^{S}$, we finally have
\begin{equation}
P(x^0,x^1,\cdots)\propto \exp\left(-\frac12 \Delta l^2\right)
\end{equation}
where
\begin{equation}\label{line}
\Delta l^2=-\frac{\partial^2 S}{\partial x^\mu \partial x^\nu}\Delta x^\mu \Delta x^\nu,
\end{equation}
is called as the metric of the Ruppeiner thermodynamic geometry (here we omit subscript $B$).

For a system composed of a black hole and its surrounding infinite environment, the black hole itself is a small subsystem of the above. Back to our focused AdS black hole, the first law of thermodynamics is $dM=TdS+VdP+\text{other terms}$. For the situation with other terms fixed, we adjust the first law of thermodynamics slightly to get
\begin{eqnarray}
dS=\frac{1}{T}dM-\frac{V}{T}dP,
\end{eqnarray}
which means that the entropy as a function of enthalpy (or mass) and thermodynamic pressure. Now we set $x^{\mu}=(M,P)$, and then the conjugate quantities corresponding to $x^{\mu}$ are $y_{\mu}=\partial S/\partial x^{\mu}=(1/T,-V/T)$. Hence the line element Eq.~(\ref{line}) becomes $\Delta l^2=-\Delta y_{\mu} \Delta x^{\mu}$. Finally, we can write the line element Eq.~(\ref{line}) as a universal form for the AdS black hole~\cite{Xu2020}
\begin{equation}\label{uline}
\Delta l^2=\frac{1}{T}\Delta T \Delta S+\frac{1}{T}\Delta V \Delta P.
\end{equation}

The phase space of the AdS black hole is $\{T, P, S, V\}$. For the theory of thermodynamics geometry, we do it in a space of generalized coordinates, like as $\{T, P\}$, $\{S, V\}$, $\{T, V\}$ and $\{S, P\}$ for the AdS black hole. We can realize that the thermodynamic curvatures obtained in these coordinate spaces are same. Therefore, we take the coordinate space $\{S,P\}$ as an example for subsequent calculation and analysis in this paper. According to Eq.~(\ref{uline}), the line element of the Ruppeiner geometry for the AdS black hole takes the form in the coordinate space $\{S,P\}$,
\begin{equation}\label{linesp}
\Delta l^2=\frac{1}{T}\left(\frac{\partial T}{\partial S}\right)_P \Delta S^2+\frac{2}{T}\left(\frac{\partial T}{\partial P}\right)_S \Delta S \Delta P+\frac{1}{T}\left(\frac{\partial V}{\partial P}\right)_S \Delta P^2,
\end{equation}
where we have used the Maxwell relation $(\partial T/\partial P)_{_S}=(\partial V/\partial S)_{_P}$ based on the first law of thermodynamics.

We use the Christoffel symbols,
\begin{equation}
\Gamma^{\alpha}_{\beta\gamma}=\frac12 g^{\mu\alpha}\left(\partial_{\gamma}g_{\mu\beta}+\partial_{\beta}g_{\mu\gamma}-\partial_{\mu}g_{\beta\gamma}\right), \label{rgc}
\end{equation}
and then write the Riemannian curvature tensor,
\begin{equation}
{R^{\alpha}}_{\beta\gamma\delta}=\partial_{\delta}\Gamma^{\alpha}_{\beta\gamma}-\partial_{\gamma}\Gamma^{\alpha}_{\beta\delta}+
\Gamma^{\mu}_{\beta\gamma}\Gamma^{\alpha}_{\mu\delta}-\Gamma^{\mu}_{\beta\delta}\Gamma^{\alpha}_{\mu\gamma}. \label{rgr}
\end{equation}
Hence, we obtain the thermodynamic (scalar) curvature,
\begin{equation}\label{curvature}
R=g^{\mu\nu}{R^{\xi}}_{\mu\xi\nu}.
\end{equation}

\section{New measure of thermal micro-behavior of the AdS black hole} \label{sec3}
To study the relationship between the interaction among black hole molecules and their thermal motion, as a preliminary and coarse-grained description, we define the following dimensionless ratio to characterize the strength relation between the interaction and the thermal motion,
\begin{eqnarray}\label{ratio}
\eta: &=&\frac{\text{Interaction}}{\text{Thermal motion}}\nonumber\\
\vspace{2mm}\nonumber\\
&\approx & \frac{\text{The magnitude of thermodynamic curvature $\times$ Planck volume}}{\text{Temperature $\times$ entropy}}.
\end{eqnarray}
Several points are explained as follows:

\begin{itemize}
  \item In usual thermodynamics, the key notion which is used to describe the phase transition is the free energy, which is supposed to measure the competition between interactions and the thermal motions. In particular, in free energy, the degree of the thermal motion is measured by the product of the temperature and entropy of the system.
  \item Firstly, through analysis, it can be seen that the temperature has the dimension of $[\text{length}]^{-1}$, the entropy has the dimension of $[\text{length}]^{2}$, and the volume has the dimension of $[\text{length}]^{3}$. Meanwhile, the thermodynamic curvature has the dimension of $[\text{length}]^{-2}$, which is consistent with the dimension of thermodynamic pressure in the natural system of units. In order to keep the same dimension as the measurement (the product of the temperature and entropy) of the thermal motion, and as an analogy with $PV$ term in ordinary thermodynamics, we think that the combination of the thermodynamic curvature and a certain volume may be a good physical quantity to describe the degree of the interaction between black hole molecules.
      Secondly, Ruppeiner~\cite{Ruppeiner2008} proposed that if we imagine the fluid broken up into pieces each of volume $v$, $|R|$ is the average number of correlated ``pixels''. We assume that each pixel occupies about one Planck volume, because Planck length is often suggested as a physical constant in quantum gravity scale. Hence we can approximately think that the product of the absolute value of thermodynamic curvature and the size of each pixel can qualitatively reflect the strength of the interaction. Again, our description here is only a preliminary exploration and there is no direct evidence for such a conjecture at present.
  \item The dimensions of the numerator and denominator in the above definition~(\ref{ratio}) are consistent. When the interaction represented by the numerator and the thermal motion represented by the denominator are in balance with each other, the ratio $\eta$ would be a fixed constant and we can always set the fixed constant to be one. Hence $\eta=1$ is special and important.
  \item If $\eta>1$, this means that the interaction among molecules is dominant in the black hole system.
  \item If $\eta<1$, this means that the thermal motion of molecules is dominant in the black hole system.
  \item If $\eta=1$, this means that the interaction among molecules and the thermal motion of molecules reach a competitive balance, and the whole system will be in a transition from the interaction state to the thermal motion state, or vice versa.
\end{itemize}
Next, we will use the newly introduced measurement to investigate the micro behaviors of several kinds of AdS black holes. In the following discussion, we often set the value of Planck volume as a unit.

\subsection{Four-dimensional charged AdS black hole}
We start with the four-dimensional charged AdS black hole and its metric is~\cite{Kubiznak2012,Spallucci2013,Niu2012}
\begin{equation}
d s^2=-f(r)dt^2+\frac{d r^2}{f(r)}+r^2(d\theta^2+\sin^2 \theta d\varphi^2),
\end{equation}
here the function $f(r)=1-2M/r+r^2/l^2+q^2/r^2$ where $M$ is the mass of the black hole, $l$ is the curvature radius of the AdS spacetime and $q$ is the total charge of the black hole. The horizon radius $r_h$ is regarded as the largest root of equation $f(r)=0$, and then the temperature of the black hole writes as
\begin{equation}
T=\frac{8P S^2+S-\pi q^2}{4S\sqrt{\pi S}},\label{temperature}
\end{equation}
where the entropy $S=\pi r_h^2$ and thermodynamic pressure $P=3/(8 \pi l^2)$. Furthermore the thermodynamic volume is $V=4\pi r_h^3/3$. Hence according to Eq.~(\ref{curvature}), we obtain the thermodynamic curvature of the four-dimensional charged AdS black hole,
\begin{equation}
R=\frac{2\pi q^2-S}{S(8P S^2+S-\pi q^2)}. \label{charge}
\end{equation}

For the sake of convenience, one usually introduces some dimensionless reduced parameters as follows~\cite{Kubiznak2012,Miao2018},
\begin{equation}
t:=\frac{T}{T_c}, \qquad s:=\frac{S}{S_c}, \qquad p:=\frac{P}{P_c}, \qquad \zeta:=\left|\frac{R}{R_c}\right|.\label{redp}
\end{equation}
where
\begin{equation}
T_c=\frac{\sqrt{6}}{18\pi q}, \qquad S_c=6\pi q^2, \qquad P_c=\frac{1}{96\pi q^2}, \qquad R_c=-\frac{1}{12\pi q^2}. \label{cv}
\end{equation}

We have the dimensionless ratio between the dimensionless measurement $ts$ of the thermal motion and the dimensionless measurement $\zeta$ of the interaction
\begin{equation}
\eta=\left|\frac{32(1-3s)}{\sqrt{s}(3ps^2+6s-1)^2}\right|.
\end{equation}

\begin{figure}
\begin{center}
\subfigure[$P/P_c=0.5$]{
\includegraphics[width=7cm]{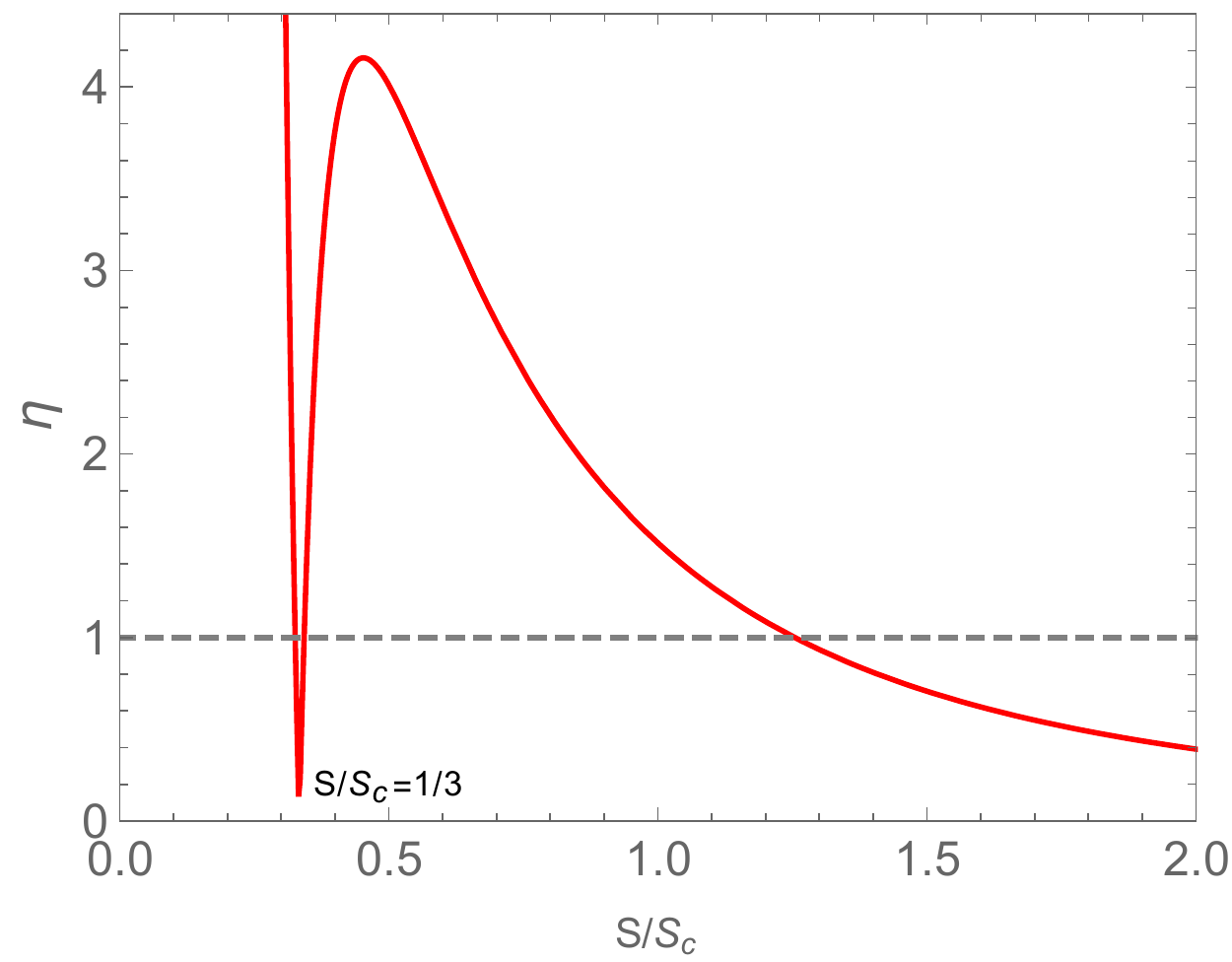}
}
\quad
\subfigure[$P/P_c=1.0$]{
\includegraphics[width=7cm]{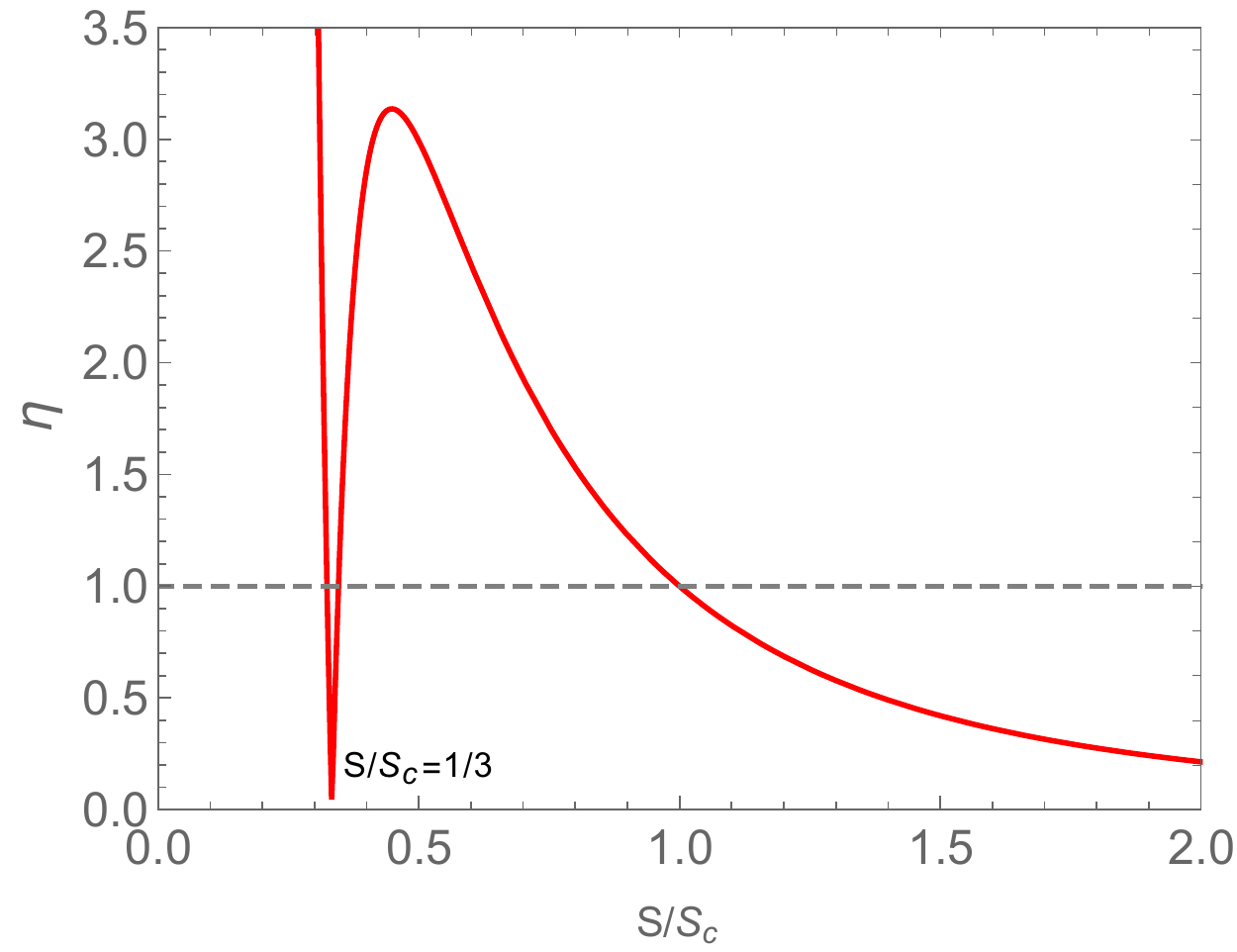}
}
\quad
\subfigure[$P/P_c=3.93$ (critical point)]{
\includegraphics[width=7cm]{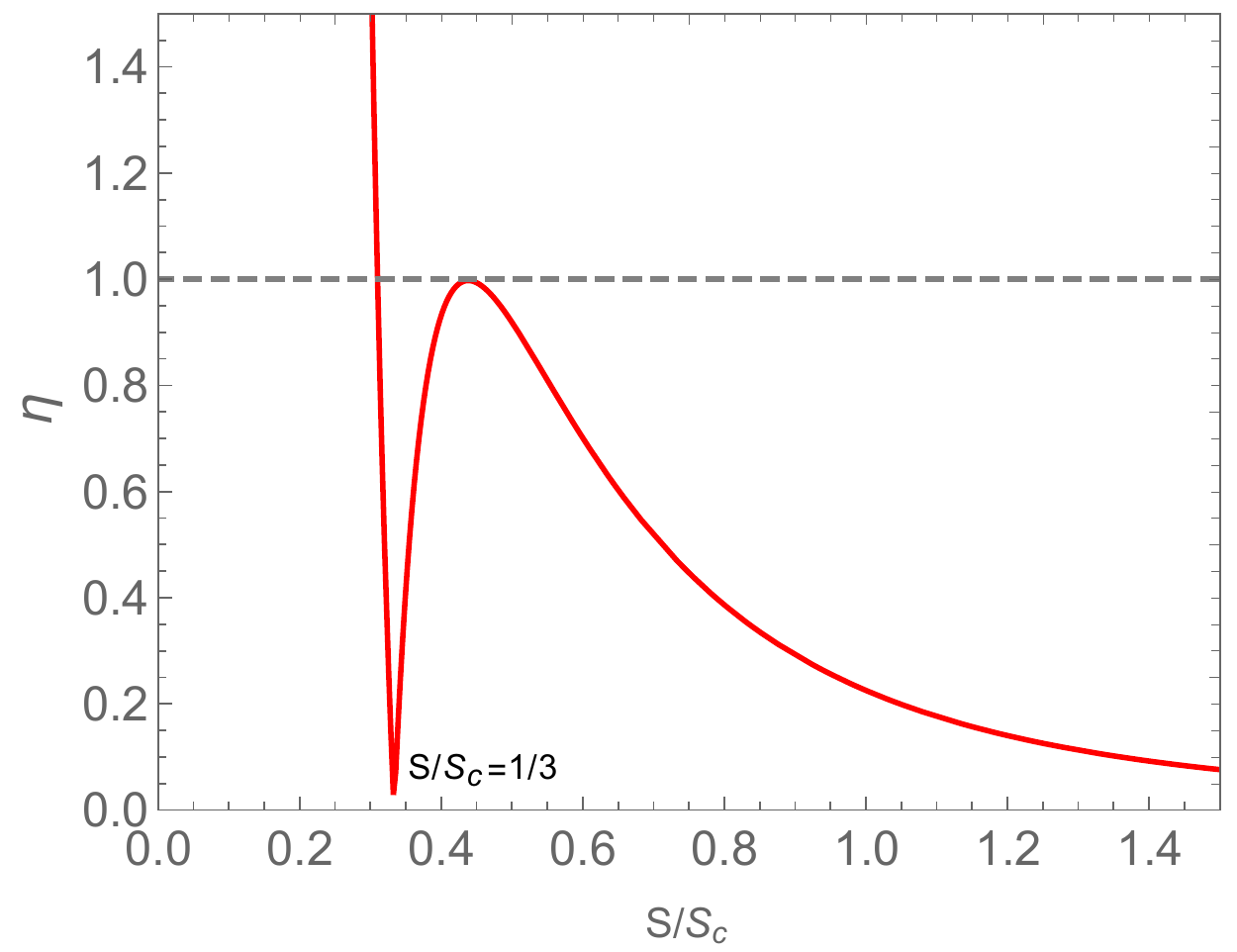}
}
\quad
\subfigure[$P/P_c=5.0$]{
\includegraphics[width=7cm]{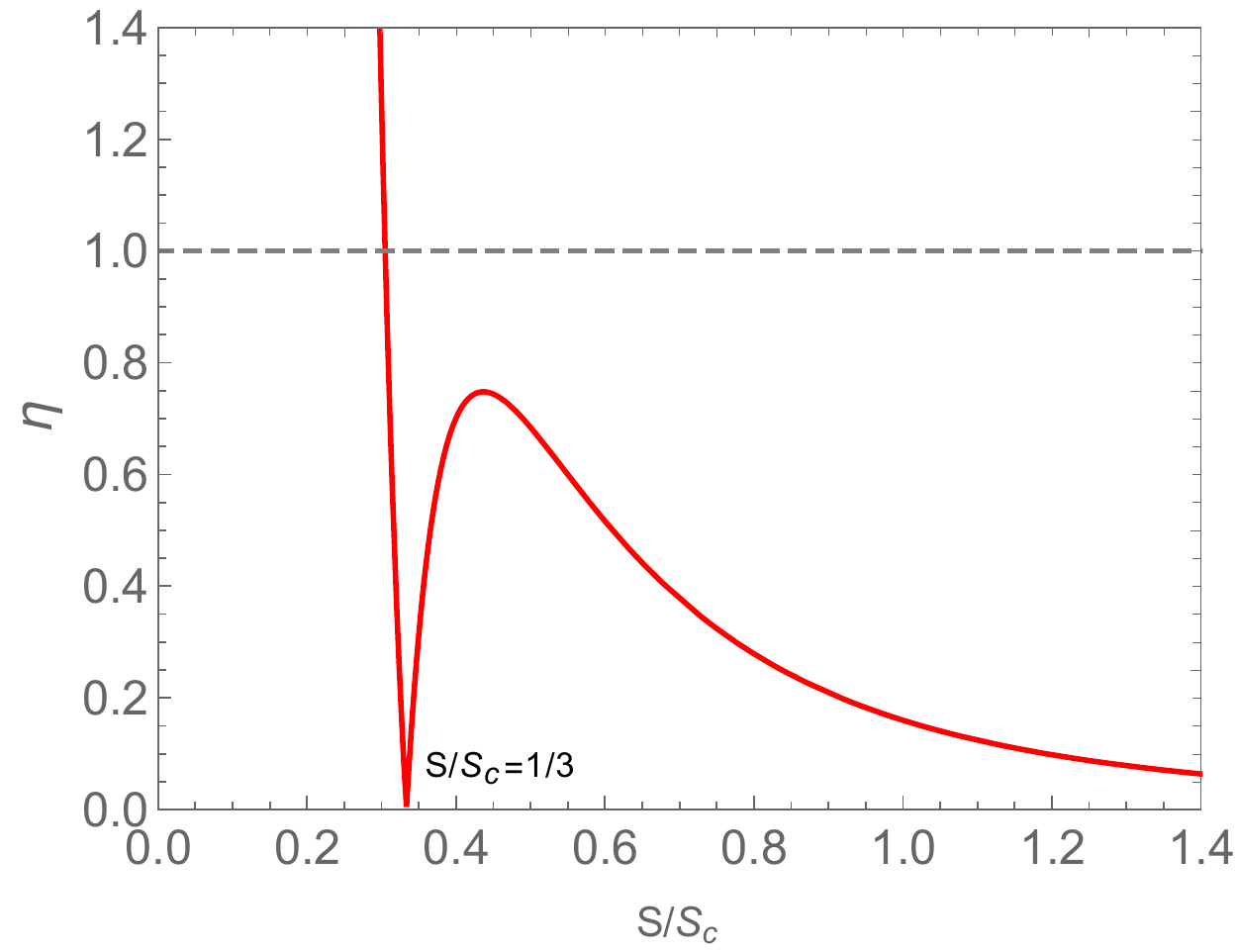}
}
\end{center}
\caption{The dimensionless ratio $\eta$ between the interaction and the thermal motion with respect to the dimensionless entropy $S/S_c$ at the different dimensionless pressure $P/P_c$ for the charged AdS black hole, respectively.}
\label{fig1}
\end{figure}
Then we plot the behaviors of the dimensionless ratio $\eta$ with respect to the dimensionless entropy $s$ at the different dimensionless pressure $p$ respectively in FIG. \ref{fig1}. Therefore, we can analyze some novel properties of thermal micro-behaviors of the charged AdS black hole:
\begin{itemize}
  \item For the charged AdS black hole system, there is indeed competition between the interaction among black hole molecules and their thermal motion.
  \item At $s=1/3$, the thermodynamic curvature equals zero. Therefore, we can divide the microscopic behaviors of the black hole into two branches. The branch-1 (B1) corresponds to the interval of $0<s<1/3$ and the branch-2 (B2) corresponds to the interval of $s>1/3$.
  \item In B1, we can see that no matter what the value of the dimensionless $p$ is, with the increasing of the dimensionless entropy $s$, the dimensionless ratio $\eta$ decreases monotonically and always has an intersection with the curve $\eta=1$. This implies that the black hole always experiences a transition from the interaction state to the thermal motion state.
  \item In B2, with the increasing of the dimensionless entropy $s$, the dimensionless ratio $\eta$ shows a trend of increasing first and then decreasing, and its intersection with the curve $\eta=1$ depends on the value of the dimensionless pressure $p$. Through numerical calculation, the new critical dimensionless pressure is $p\approx 3.93$. When $0< p < 3.93$, there are two intersections. The first intersection represents the transition from the thermal motion state to the interaction state, and the second one implies the transition from the interaction state to the thermal motion state. When $p\geq 3.93$, the two intersections merge and disappear resulting that the black hole will always be in the thermal motion state.
\end{itemize}

\subsection{Four-dimensional Schwarzschild-AdS black hole}
The Schwarzschild-AdS black hole is a special case of the charged AdS black hole with $q=0$. Its temperature and the thermodynamic curvature read as~\cite{Xu2020}
\begin{equation}
T=\frac{8P S+1}{4\sqrt{\pi S}}, \qquad R=-\frac{1}{8P S^2+S}.
\end{equation}
We introduce a dimensionless quantity $u=8PS$, then the temperature, entropy and the thermodynamic curvature can also be expressed as the rescaled temperature $t$, entropy $s$ and thermodynamic curvature $\zeta$
\begin{equation}
t=\frac{T}{\sqrt{8\pi P}}=\frac{u+1}{4\pi\sqrt{u}}, \qquad s=8\pi PS=\pi u, \qquad \zeta=\left|\frac{R}{8P}\right|=\frac{1}{u(u+1)}.
\end{equation}
Hence the dimensionless ratio $\eta$ reads as
\begin{equation}
\eta=\frac{4}{u^{3/2}(u+1)}.
\end{equation}

\begin{figure}
 \begin{center}
 \includegraphics[width=90mm]{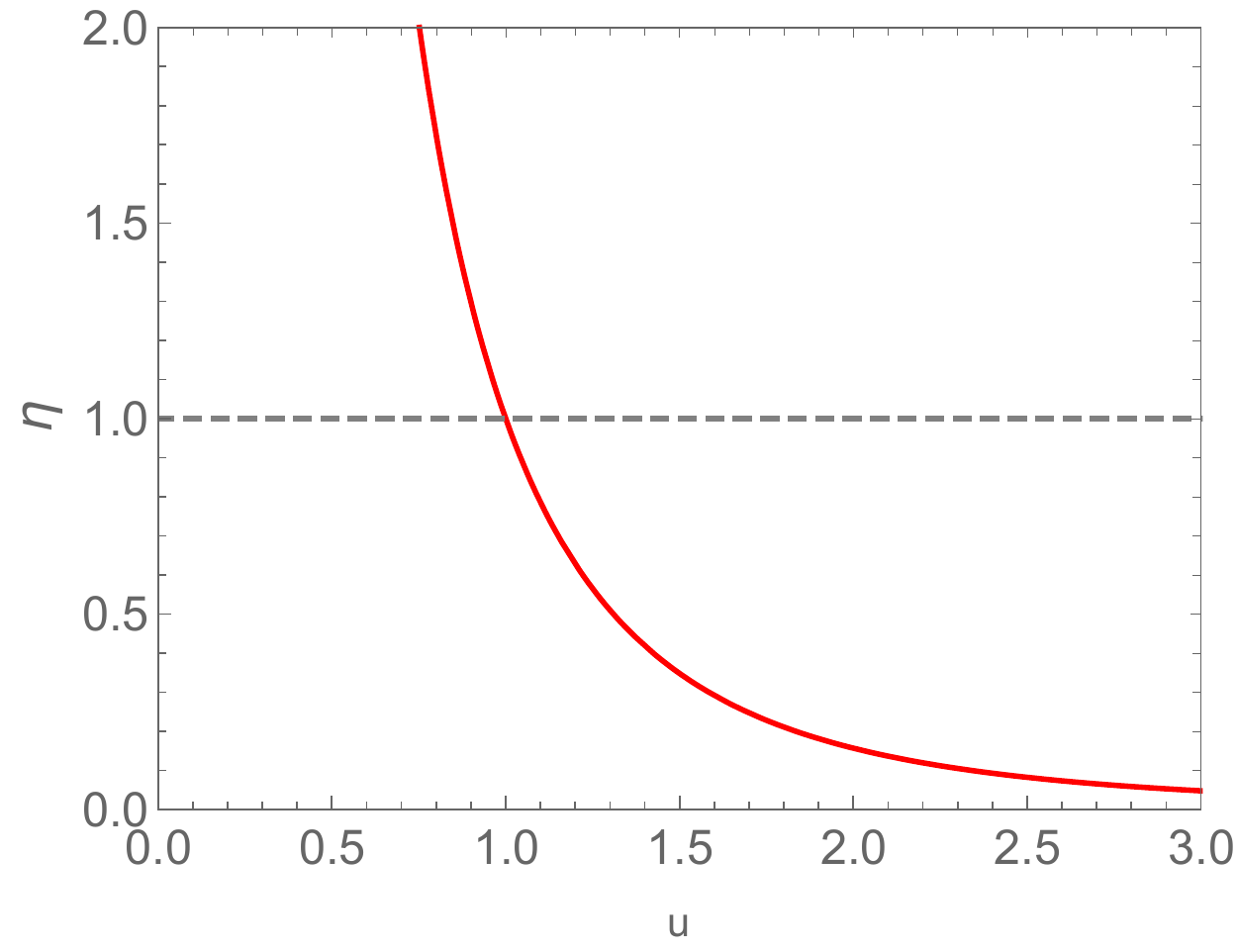}
 \end{center}
 \caption{The dimensionless ratio $\eta$ between the interaction and the thermal motion with respect to the quantity $u$ for the Schwarzschild-AdS black hole.}
 \label{fig2}
 \end{figure}

We plot the behavior of the dimensionless ratio $\eta$ between the thermal motion and the interaction with respect to the quantity $u$ in FIG. \ref{fig2}. We can clearly see that with the increasing of $u$, the dimensionless ratio $\eta$ decreases monotonically and always has an intersection with the curve $\eta=1$ at $u=1$. This implies that the Schwarzschild-AdS black hole always experiences a transition from the interaction state to the thermal motion state.

\subsection{Five-dimensional Gauss-Bonnet AdS black hole}
The metric of the Gauss-Bonnet AdS black hole in $d$ dimensions is \cite{Cai2002}
\begin{equation}
ds^2=-f(r)dt^2+\frac{1}{f(r)}dr^2+r^2 d\Omega^2,
\end{equation}
and
\begin{equation*}
f(r)=1+\frac{r^2}{2\alpha_0}\left(1-\sqrt{1+\frac{64\pi\alpha_0 {\cal M}}{(d-2)r^{d-1}\Sigma}-\frac{64\pi\alpha_0 P}{(d-1)(d-2)}}\right),
\end{equation*}
where $\mathrm{d}\Omega^2$ is the square of line element on a $(d-2)$-dimensional maximally symmetric Einstein manifold with volume $\Sigma$. The black hole mass is $\cal M$ and the pressure is $P=(d-1)(d-2)/(16\pi l^2)$. The auxiliary symbol $\alpha_0$ is related to the Gauss-Bonnet coefficient $\alpha_{_{\mathrm{GB}}}$ via $\alpha_0= (d-3)(d-4)\alpha_{_{\mathrm{GB}}}$ in order to avoid the verboseness.

When $d=5$, its temperature and entropy takes the following form in terms of the horizon radius $r_h$  \cite{Cai2002,Xu2017}
\begin{equation}
T=\frac{8\pi P r_h^3+3r_h}{6\pi(r_h^2+2\alpha_0)}, \qquad S=\frac{\pi^2 r_h(r_h^2+6\alpha_0)}{2}.
\end{equation}

Meanwhile the thermodynamic curvature for the five-dimensional Gauss-Bonnet AdS black hole in terms of the horizon radius $r_h$ is  \cite{Miao2018b}
\begin{equation}
R=-\frac{4}{\pi^2 r_h(r_h^2+2\alpha_0)(8\pi P r_h^2+3)}.
\end{equation}

\begin{figure}
 \begin{center}
 \includegraphics[width=90mm]{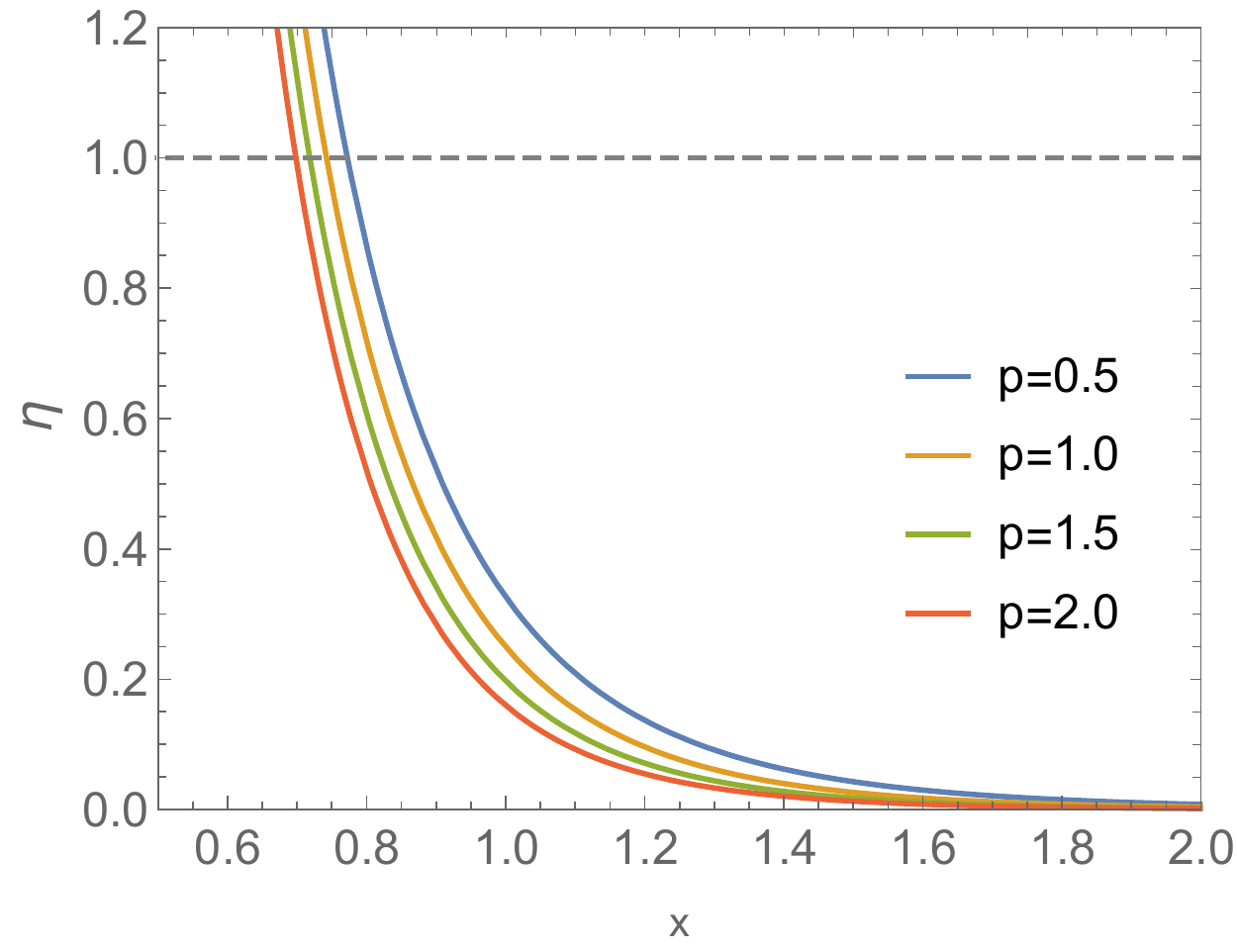}
 \end{center}
 \caption{The dimensionless ratio $\eta$ between the interaction and the thermal motion with respect to the quantity $x$ for the five-dimensional Gauss-Bonnet AdS black hole.}
 \label{fig3}
\end{figure}

Similarly, the dimensionless temperature $t$, entropy $s$ and thermodynamic curvature $\zeta$ can be written as
\begin{equation}
t=\frac{p x^3+3x}{3x^2+1}, \qquad s=\frac{x^3+x}{2}, \qquad \zeta=\frac{4}{(3x^2+1)(p x^3+3x)},
\end{equation}
with $t=T/T_c$, $s=S/S_c$, $p=P/P_c$, $\zeta=|R/R_c|$ and $x=r_h/r_c$ where  \cite{Cai2002,Xu2017,Miao2018b}
\begin{eqnarray*}
&&T_c=\frac{1}{2\pi\sqrt{6\alpha_0}}, \qquad r_c=\sqrt{6\alpha_0}, \qquad S_c=6 \pi^2 \alpha_0 \sqrt{6\alpha_0}, \\
&&P_c=\frac{1}{48\pi \alpha_0}, \qquad R_c=-\frac{1}{8\pi^2 \alpha_0 \sqrt{6\alpha_0}}.
\end{eqnarray*}
Hence the dimensionless ratio between the thermal motion and the interaction is
\begin{equation}
\eta=\frac{8}{(p x^3+3x)^2(x^3+x)}.
\end{equation}
The behavior of the ratio $\eta$ with respect to the $x$ in different dimensionless pressure $p$ in FIG. \ref{fig3}. Clearly, with the increasing of $x$, the dimensionless ratio $\eta$ decreases monotonically and always has an intersection with the curve $\eta=1$. This implies that the five-dimensional Gauss-Bonnet AdS black hole always experiences a transition between the interaction state and the thermal motion state.

\section{Summary and Discussion}\label{sec4}
In the absence of a complete theory of quantum gravity, thermodynamic geometry can be said to be one of the effective means to study the microscopic behavior of black holes from the perspective of thermodynamics. In~\cite{Xu2019}, by introducing a new generalized coordinate, we preliminarily solve the puzzle whether the Reissner-Nordstr\"{o}m black hole is an interacting system or not. Meanwhile, in the analysis of the microscopic behavior of the Reissner-Nordstr\"{o}m black hole, we give a signal about the relation between the thermal motion of black hole molecules and their interaction. In~\cite{Xu2020}, we propose a scheme to solve the problem of the singularity of thermodynamic geometry for static spherically symmetric AdS black holes. We point out that enthalpy is the basic thermodynamic potential of thermodynamic geometry for such black holes, from which we can obtain the thermodynamic curvature with good behavior. The scheme of the paper~\cite{Xu2020} is also an important basis for our current work. This method is different from that in the literature~\cite{Wei2019a}, where the normalized thermodynamic curvature has been introduced to cure the divergence of the thermodynamic curvature, and the internal energy is regarded as the basic thermodynamic potential. In~\cite{Xu2020b}, with the help of the scheme proposed in~\cite{Xu2020}, we analyze the behavior of the thermodynamic curvature for the BTZ black hole, and give a diagnosis for the discrimination of different thermodynamic calculation schemes from the point of view of the thermodynamics geometry of the BTZ black hole.

Now in this paper, under the view of the black hole molecule, we introduce a dimensionless ratio~(\ref{ratio}) to characterize the relation between the interaction and the thermal motion of the AdS black holes. Through a preliminary and coarse-grained description, we find that there is indeed competition between the interaction among black hole molecules and their thermal motion.

\begin{itemize}
  \item For the charged AdS black hole, the microscopic behaviors of the black hole can be divided into two branches (B1 and B2). In B1, the black hole always experiences a transition from the interaction state to the thermal motion state. In B2, the transition point between the interaction state and the thermal motion state depends on the value of the dimensionless pressure $p$. Below the critical dimensionless pressure $p=3.93$, there will be two transitions, while above the critical dimensionless pressure, the transition disappears and the black hole is always in the thermal motion state.
  \item For the Schwarzschild-AdS black hole, there is always a transition between the interaction state and the thermal motion state. Meanwhile, the transition point exactly satisfies $8PS=1$.
  \item For the five-dimensional Gauss-Bonnet AdS black hole, it always experiences a transition between the interaction state and the thermal motion state.
\end{itemize}

In addition, our current approach can naturally be extended to other types of black holes, like (charged) Kerr (-AdS) black holes, by which we can analyze some new thermal micro information of various black holes.

\section*{Acknowledgments}
The financial supports from National Natural Science Foundation of China (Grant No.11947208, Grant No.11947301), China Postdoctoral Science Foundation (Grant No. 2020M673460), Major Basic Research Program of Natural Science of Shaanxi Province (Grant No.2017KCT-12), Scientific Research Program Funded by Shaanxi Provincial Education Department (Program No.18JK0771) are gratefully acknowledged. This research is also particularly supported by The Double First-class University Construction Project of Northwest University. The authors would like to thank the anonymous reviewers for the helpful comments that indeed greatly improve this work.


\begin{thebibliography}{99}
\bibitem{Hawking1975}S. Hawking, Particle creation by black holes, Commun. Math. Phys. 43 (1975) 199; Erratum ibid. 46 (1976) 206.

\bibitem{Bekenstein1973}J.D. Bekenstein, Black holes and entropy, Phys. Rev. D 7 (1973) 2333.

\bibitem{Bardeen1973}J.M. Bardeen, B. Carter, and S. Hawking, The four laws of black hole mechanics, Commun. Math. Phys. 31 (1973) 161.

\bibitem{Hawking1983}S. Hawking and D.N. Page, Thermodynamics of black holes in anti-de Sitter space, Commun. Math. Phys. 87 (1983) 577.

\bibitem{Chamblin1999}A. Chamblin, R. Emparan, C.V. Johnson, and R.C. Myers, Holography, thermodynamics and fluctuations of charged AdS black holes, Phys. Rev. D 60 (1999) 104026.

\bibitem{Wald2001}R.M. Wald, The thermodynamics of black holes, Living Rev. Rel. 4 (2001) 6.

\bibitem{Padmanabhan2010}T. Padmanabhan, Thermodynamical aspects of gravity: New insights, Rep. Prog. Phys. 73 (2010) 046901.

\bibitem{Carlip2014}S. Carlip, Black hole thermodynamics, Int. J. Mod. Phys. D 23 (2014) 1430023.

\bibitem{Kastor2009}D. Kastor, S. Ray, and J. Traschen, Enthalpy and the mechanics of AdS black holes, Class. Quant. Grav. 26 (2009) 195011.

\bibitem{Dolan2011a}B.P. Dolan, The cosmological constant and black-hole thermodynamic potentials, Class. Quant. Grav. 28 (2011) 125020.

\bibitem{Kubiznak2012}D. Kubiznak and R.B. Mann, $P-V$ criticality of charged AdS black holes, JHEP 07 (2012) 033.

\bibitem{Spallucci2013}E. Spallucci and A. Smailagic, Maxwell’s equal area law for charged anti-de Sitter black holes, Phys. Lett. B 723 (2013) 436.

\bibitem{Kubiznak2017}D. Kubiznak, R.B. Mann, and M. Teo, Black hole chemistry: thermodynamics with Lambda, Class. Quantum Grav. 34 (2017) 063001.

\bibitem{Ruppeiner1995}G. Ruppeiner, Riemannian geometry in thermodynamic fluctuation theory, Rev. Mod. Phys. 67 (1995) 605; Erratum ibid. 68 (1996) 313.

\bibitem{Wei2015}S.-W. Wei and Y.-X. Liu, Insight into the microscopic structure of an AdS black hole from a thermodynamical phase transition, Phys. Rev. Lett. 115 (2015) 111302; Erratum ibid. 116 (2016) 169903.

\bibitem{Cai1999}R.-G. Cai and J. H. Cho, Thermodynamic curvature of the BTZ black hole, Phys. Rev. D 60 (1999) 067502.

\bibitem{Mirza2007}B. Mirza, M. Zamani-Nasab, Ruppeiner geometry of RN black holes: flat or curved?, JHEP 0706 (2007) 059.

\bibitem{Dehyadegari2017}A. Dehyadegari, A. Sheykhi, and A. Montakhab, Critical behavior and microscopic structure of charged AdS black holes via an alternative phase space, Phys. Lett. B 768 (2017) 235.

\bibitem{Miao2018}Y.-G. Miao and Z.-M. Xu, Thermal molecular potential among micromolecules in charged AdS black holes, Phys. Rev. D 98 (2018) 044001.

\bibitem{Zangeneh2018}M.K. Zangeneh, A. Dehyadegari, A. Sheykhi, and R. B. Mann, Microscopic origin of black hole reentrant phase transitions, Phys. Rev. D 97 (2018) 084054.

\bibitem{Yazdikarimi2019}H. Yazdikarimi, A. Sheykhi, and Z. Dayyani, Critical behavior of Gauss-Bonnet black holes via an alternative phase space, Phys. Rev. D 99 (2019) 124017.

\bibitem{Guo2019}X.-Y. Guo, H.-F. Li, L.-C. Zhang and R. Zhao, Microstructure and continuous phase transition of a Reissner-Nordstrom-AdS black hole, Phys. Rev. D 100 (2019) 064036.

\bibitem{Miao2019}Y.-G. Miao and Z.-M. Xu, Interaction potential and thermo-correction to the equation of state for thermally stable Schwarzschild anti-de Sitter black holes, Sci. China-Phys. Mech. Astron. 62 (2019) 010412.

\bibitem{Xu2019}Z.-M. Xu, B. Wu, and W.-L. Yang, The fine micro-thermal structures for the Reissner-Nordstr\"{o}m black hole, Chinese Phys. C 44 (2020) 095106.

\bibitem{Xu2020}Z.-M. Xu, B. Wu, and W.-L. Yang, Ruppeiner thermodynamic geometry for the Schwarzschild AdS black hole, Phys. Rev. D 101 (2020) 024018.

\bibitem{Xu2020b}Z.-M. Xu, B. Wu, and W.-L. Yang, Diagnosis inspired by the thermodynamic geometry for different thermodynamic schemes of the charged BTZ black hole, arXiv:2002.00117 [gr-qc].

\bibitem{Ghosh2020}A. Ghosh and C. Bhamidipati, Thermodynamic geometry for charged Gauss-Bonnet black holes in AdS spacetimes, Phys. Rev. D 101 (2020) 046005.

\bibitem{Shen2007}J.-Y. Shen, R.-G. Cai, B. Wang and R.-K. Su, Thermodynamic geometry and critical behavior of black holes, Int. J. Mod. Phys. A 22 (2007) 11.

\bibitem{Liu2010}H.-S. Liu, H. Lu, M.-X Luo, and K.-N. Shao, Thermodynamical metrics and black hole phase transitions, JHEP 12 (2010) 054.

\bibitem{Niu2012}C. Niu, Y. Tian, and X.-N. Wu, Critical phenomena and thermodynamic geometry of Reissner-Nordstrm-anti-de Sitter black holes, Phys. Rev. D 85 (2012) 024017.

\bibitem{Wang2020}P. Wang, H.-W. Wu, and H.-T. Yang, Thermodynamic geometry of AdS black holes and black holes in a cavity, Eur. Phys. J. C 80 (2020) 216.

\bibitem{Mansoori2015}S.A.H. Mansoori, B. Mirza and M. Fazel, Hessian matrix, specific heats, Nambu brackets, and thermodynamic geometry, J. High Energy Phys. 04 (2015) 115.

\bibitem{Mansoori2019}S.A.H. Mansoori and B. Mirza, Geometrothermodynamics as a singular conformal thermodynamic geometry, Phys. Lett. B 799 (2019) 135040.

\bibitem{Quevedo2008}H. Quevedo, Geometrothermodynamics of black holes, Gen. Rel. Grav. 40 (2008) 971.

\bibitem{Bhattacharya2017}K. Bhattacharya and B.R. Majhi, Thermogeometric description of the van der Waals like phase transition in AdS black holes, Phys. Rev. D 95 (2017) 104024.

\bibitem{Bhattacharya2020}K. Bhattacharya and B.R. Majhi, Thermogeometric study of van der Waals like phase transition in black holes: an alternative approach, Phys. Lett. B 802 (2020) 135224.

\bibitem{Zhang2015}J.-L. Zhang, R.-G. Cai, and H.-W. Yu, Phase transition and thermodynamical geometry of Reissner-Nordstr\"{o}m-AdS black holes in extended phase space, Phys. Rev. D 91 (2015) 044028.

\bibitem{Ruppeiner2010}G. Ruppeiner, Thermodynamic curvature measures interactions, Am. J. Phys. 78 (2010) 1170.

\bibitem{Ruppeiner2014}G. Ruppeiner, Thermodynamic curvature and black holes, in Breaking of Supersymmetry and Ultraviolet Divergences in Extended Supergravity, edited by S. Bellucci, Springer Proceedings in Physics, Vol. 153 (Springer, New York, 2014), p. 179.

\bibitem{Ruppeiner2008}G. Ruppeiner, Thermodynamic curvature and phase transitions in Kerr-Newman black holes, Phys. Rev. D 78 (2008) 024016.

\bibitem{Wei2019a}S.-W. Wei, Y.-X. Liu, and R.B. Mann, Repulsive interactions and universal properties of charged anti-de Sitter black hole microstructures, Phys. Rev. Lett. 123 (2019) 071103.

\bibitem{Wei2019b}S.-W. Wei, Y.-X. Liu, and R.B. Mann, Ruppeiner geometry, phase transitions, and the microstructure of charged AdS black holes, Phys. Rev. D 100 (2019) 124033.

\bibitem{Wei2019c}S.-W. Wei and Y.-X. Liu, Intriguing microstructures of five-dimensional neutral Gauss-Bonnet AdS black hole, Phys. Lett. B 803 (2020) 135287.

\bibitem{Cai2002}R.-G. Cai, Gauss-Bonnet black holes in AdS spaces, Phys. Rev. D 65 (2002) 084014.

\bibitem{Xu2017}H. Xu and Z.-M. Xu, Maxwell’s equal area law for Lovelock thermodynamics, Int. J. Mod. Phys. D 26 (2017) 1750037.

\bibitem{Miao2018b}Y.-G. Miao and Z.-M. Xu, Parametric phase transition for a Gauss-Bonnet AdS black hole, Phys. Rev. D 98 (2018) 084051.


\end{thebibliography}
\end{document}